\documentclass[12pt,a4]{article}
\usepackage{epsfig}
\usepackage{multirow}

\def\Title#1{\begin{center} {\Large {\bf #1} } \end{center}}
\def\Bs{${\mathrm{B}}^{\mathrm{0}}_{\mathrm{s}}$}
\def\Bsbar{${\bar{\mathrm{B}}}^{\mathrm{0}}_{\mathrm{s}}$}
\def\B{${\mathrm{B}}^{\mathrm{0}}_{\mathrm{d}}$}
\def\Bbar{${\bar{\mathrm{B}}}^{\mathrm{0}}_{\mathrm{d}}$}
\def\dms{$\Delta m_{\mathrm{s}}$}
\def\dmd{$\Delta m_{\mathrm{d}}$}
\begin{document}

\Title{New Results on \Bs~Mixing from LEP}

\bigskip\bigskip

\begin{center}  

{\it Stephen R. Armstrong\index{Armstrong, S.R.}\\
CERN EP Division\\
CH-1211 Geneva 23, SWITZERLAND}
\bigskip
\begin{center}
{\it Contribution to Flavour Physics and CP Violation \\
16-18 May 2002 Philadelphia, Pennsylvania, USA}
\end{center}
\bigskip
\end{center}
\section{Introduction}

A prime goal of contemporary heavy flavour physics is the observation
of \Bs-\Bsbar~oscillations and determination of the mass
difference \dms~to which the oscillation frequency is proportional.
With the already well-measured quantity \dmd~from studies of
\B-\Bbar~oscillations~\cite{PDG}, this would permit the extraction of
the ratio of the CKM $V_{\mathrm{ts}}$ and $V_{\mathrm{td}}$ matrix
elements 
\begin{equation}
\frac{\Delta m_{\mathrm{s}}}{\Delta m_{\mathrm{d}}}=
\frac{m_{\mathrm{B}_{\mathrm{s}}}}{m_{\mathrm{B}_{\mathrm{d}}}}
\frac{\left|V_{\mathrm{ts}}\right|^2}{\left|V_{\mathrm{td}}\right|^2}
\xi^2, \,\,\,\,\,\,
\xi^2=
\frac{F^2_{\mathrm{B}_{\mathrm{s}}}B_{\mathrm{B}_{\mathrm{s}}}}
{F^2_{\mathrm{B}_{\mathrm{d}}}B_{\mathrm{B}_{\mathrm{d}}}}.
\end{equation}
The theoretical uncertainties of roughly 10\% are embedded in
the $\xi^2$ parameter, the ratio of the decay constants and
bag parameters of the \Bs~and \B~mesons.  The phenomenological implication of
\Bs-\Bsbar~oscillations is a proper time-dependent asymmetry in the probability to
observe a {\it mixed} decay ({\it i.e.}, \mbox{\Bs$\rightarrow$\Bsbar$\rightarrow X$})
compared to an {\it unmixed} decay 
({\it i.e.}, \mbox{\Bs$\rightarrow$\Bs$\rightarrow X^{\prime}$}).
These probabilities are given as
\begin{equation}
\begin{array}{c}
\mathcal{P}_{\mathrm{mixed}} (t) = \Gamma_{\mathrm{s}} 
\frac{\mathrm{e}^{-\Gamma_{\mathrm{s}} t}}{2}
[1-\cos(\Delta m_{\mathrm{s}}t)],
\,\,\,\,\,\,
\mathcal{P}_{\mathrm{unmixed}} (t) = \Gamma_{\mathrm{s}} 
\frac{\mathrm{e}^{-\Gamma_{\mathrm{s}} t}}{2}
[1+\cos(\Delta m_{\mathrm{s}}t)],
\end{array}
\label{eqn}
\end{equation}
assuming CP conservation and small lifetime differences.  The
challenge to experiments in measuring a value for \dms~is to
determine if the \Bs~meson decay is a mixed or unmixed one, and
to measure the proper time associated to it.

The LEP experiments ALEPH~\cite{ALEPH}, DELPHI~\cite{DELPHI,
  DELPHIhad}, and OPAL~\cite{OPALDsl} have investigated
\Bs-\Bsbar~oscillations as have SLD~\cite{SLD} and CDF~\cite{CDF}.
This paper focuses on new ({\it i.e.}, released in early 2002) results
from ALEPH~\cite{ALEPH} which motivate new LEP results on \Bs-\Bsbar~oscillations
and briefly reviews the results from DELPHI and OPAL.

\section{Experimental Strategy}
\begin{figure}[htb]
\begin{center}
\epsfig{file=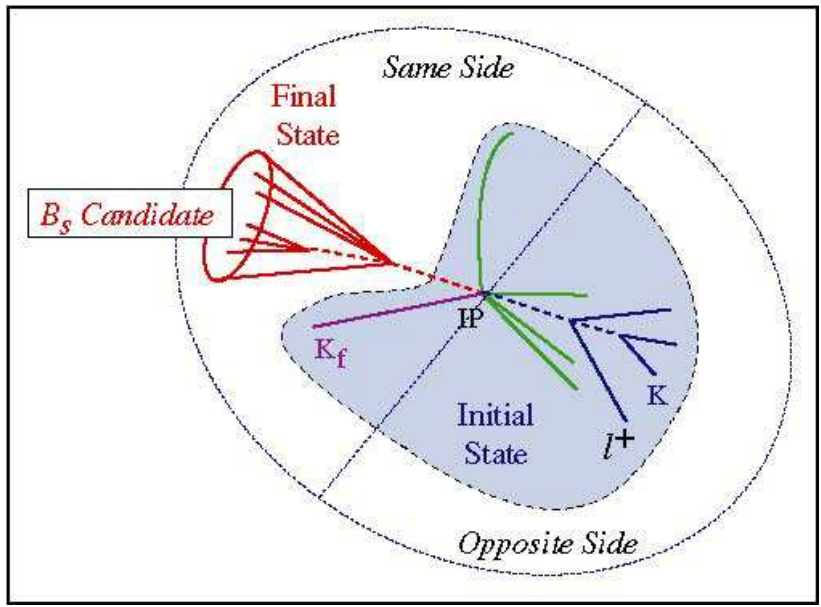,height=8cm}
\caption{{\footnotesize A diagram illustrating components of the experimental strategy.
    The event is divided into two hemispheres with respect to the
    thrust axis; the hemisphere containing the \Bs~candidate is
    referred to as the {\it Same Side} while the other is the {\it
      Opposite Side}.}}
\label{fig:strategy}
\end{center}
\end{figure}

At LEP, \Bs~mesons are produced from hadronic decays of the Z~boson
(e$^+$e$^-\rightarrow$Z$\rightarrow\mathrm{b}\overline{\mathrm{b}}$).
The boosted b hadrons result in a characteristic displaced vertex
topology relative to the interaction point, forming the basis of
most heavy flavour physics analyses.  The experimental strategy
common to all analyses studying \Bs-\Bsbar~oscillations can be grouped
into four categories discussed below.
Figure~\ref{fig:strategy} illustrates components of this strategy.
\begin{itemize}
\item{\bf{\Bs~Selection and Event Purity Determination}}\\
  Candidate \Bs~events are selected from LEP data
  collected between 1991 and 1995 (roughly 4~million
  hadronic Z~decays per experiment).\footnote{In one case, the ALEPH
    fully exclusive \Bs~selection, Z~peak calibration data from LEP2
    are included boosting the sample by about 400\,000 hadronic
    events.}  
  Table~\ref{tab:sum} summarizes different selections used by the LEP
  experiments.  Figures~\ref{fig:fullexcl}~and~\ref{fig:dsl} show
  invariant mass distributions obtained from the ALEPH fully exclusive
  and semi-exclusive analyses.
A selection-dependent event-by-event purity improves the statistical
power of the event sample~\cite{ALEPH}.  A 
probability for each candidate to originate from signal and background 
components is used in the oscillation fit described below.
\begin{figure}[htb]
\begin{center}
\epsfig{file=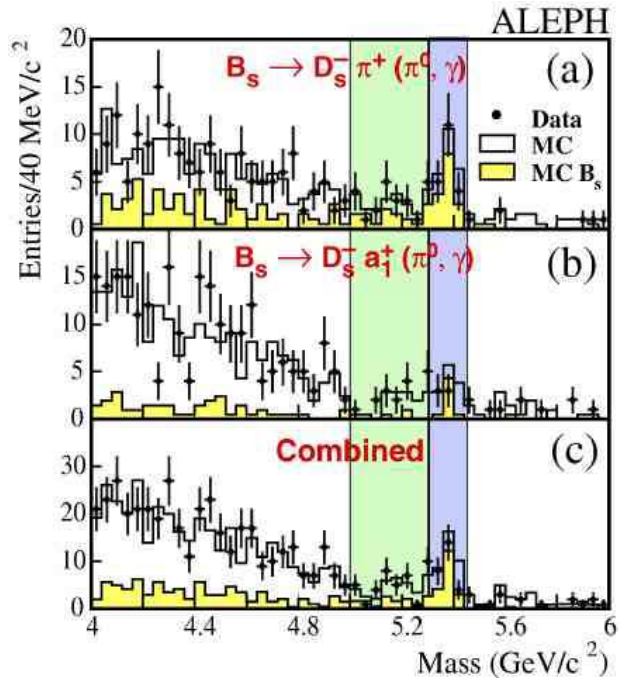,height=9cm}
\caption{{\footnotesize Invariant mass distributions for the reconstructed
  \Bs~candidates in the ALEPH fully-exclusive selection~\cite{ALEPH}.
  Data (dots with error bars) and the simulation (histograms) are
  shown: a)~the
  D$_{\mathrm{s}}^- \pi^+(\pi^0,\gamma)$ channel (a total of 44 events); b)~the
  D$_{\mathrm{s}}^-\mathrm{a}_{\mathrm{1}}^+(\pi^0,\gamma)$ channel (a total of
  36 events);
  and c)~the sum of the two.  The lightly and darkly shaded
  vertical bands show the satellite regions and the main peak area
  defining the mass selection windows.
  }}
\label{fig:fullexcl}
\end{center}
\end{figure}
\begin{figure}[htb]
\begin{center}
\epsfig{file=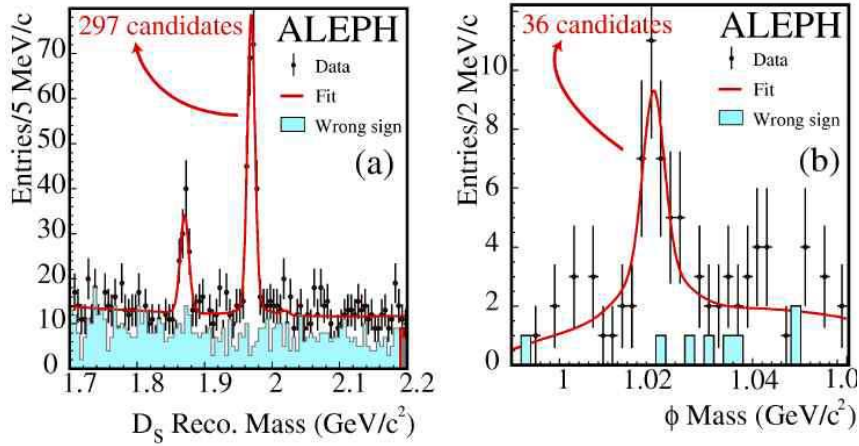,height=6cm}
\caption{{\footnotesize Invariant mass distributions for the ALEPH semi-exclusive
    selection~\cite{ALEPH} showing a)~the selected
    D$^{-}_{\mathrm{s}}$ candidates with D$^{-}_{\mathrm{s}}\ell^+$
    combinations for hadronic D$^{-}_{\mathrm{s}}$ decays, and b) the
    selected $\phi$ candidates for semileptonic D$^{-}_{\mathrm{s}}$
    decays in data (dots with error bars). }}
\label{fig:dsl}
\end{center}
\end{figure}
\begin{table}[h]
\footnotesize
\begin{center}
\begin{tabular}{|l|c|c|c|c|c|}
\hline\hline  
Selection & Decay & Sample  & Purity & LEP        &   \dms~Limit\\
          & Modes & (events)    & \%     & Experiment & obs. (exp.) ps$^{-1}$\\
\hline
Fully Exclusive  & 
\Bs$\rightarrow\mathrm{D}_{\mathrm{s}}^{-}(\pi^{+}\,\,{\mathrm{or}}\,\,{\mathrm{a}}_{1}^{+})$ &  $50\,-\,80$ & $50\,-\,80$ & ALEPH~\cite{ALEPH}      & 2.5 (0.4)\\ \cline{5-6}
                 & \Bs$\rightarrow{\overline{\mathrm{D}}}^{0}{\mathrm{K}}^{-}(\pi^{+}\,\,{\mathrm{or}}\,\,{\mathrm{a}}_{1}^{+})$ &              &             & DELPHI~\cite{DELPHIhad} &$\dagger$4.0 (3.2)$\dagger$\\
\hline
Semi-Exclusive   &  \Bs$\rightarrow{\mathrm{D}}_{\mathrm{s}}^{\mathrm{(*)-}}\ell^+\nu_\ell$ & $10^2\,-\,10^3$  & $40\,-\,60$ & ALEPH~\cite{ALEPH} & 7.2 (7.5) \\ \cline{5-6}
                 & $\dagger\,$\Bs$\rightarrow\mathrm{D}_{\mathrm{s}}^{\mathrm{(*)-}}h^{+}\,\,\dagger$ & & & DELPHI~\cite{DELPHI, DELPHIhad} & 7.4 (8.1)\\ \cline{5-6}
                 & & & & OPAL~\cite{OPALDsl} & 1.0 (4.1) \\
\hline
Semi-Inclusive   &  \Bs$\rightarrow\ell^+\nu_\ell + X$ & $10^4\,-\,10^5$ & $10\,-\,20$ & ALEPH~\cite{ALEPH} & 11.4 (14.0) \\ \cline{5-6}
                 &  & & & DELPHI~\cite{DELPHIfull} & 2.0 (7.8)\\ \cline{5-6}
                 &  & & & OPAL~\cite{OPALincl} & 5.2 (7.0) \\
\hline
Fully Inclusive  & \Bs$\rightarrow X$ & $5\times 10^5$ & 10 & DELPHI~\cite{DELPHIfull} & 1.2 (4.9)\\
\hline\hline
\end{tabular}
\caption{{\footnotesize A summary of the LEP experiments' \Bs-\Bsbar~oscillation selections, their characteristics, and the resulting 95\% C.L. lower limit on \dms.  DELPHI combine their Fully Exclusive and Semi-Exclusive D$_{\mathrm{s}}^{-} h^{+}$ results providing only a combined exclusive result, denoted by the $\dagger$.}}
\label{tab:sum}
\end{center}
\end{table}

\item{{\bf Tagging the Initial and Final States}}\\
  A determination of the anti-particle/particle state of the
  \Bs~candidate at its production (initial) and decay (final) is the
  key component of the analyses.  The Final State tagging depends upon
  the selection.  For fully exclusive decays, no ambiguity exists as
  all decay products are known.  For the inclusive analyses with
  semileptonic \Bs~decays, the charge of the lepton is used accounting
  for the non-zero mistag associated with cascade decays
  $\mathrm{b}\rightarrow\mathrm{c}\rightarrow\ell$.  Initial State
  tagging is more complicated: information from both the Same and
  Opposite Sides may be used as the flavour of the b hadron in the
  Opposite Side is anti-correlated with that of the \Bs~at production.
  In each case, a variety of discriminants are used.  The Opposite
  Side tag may rely upon jet charges, primary and secondary vertex
  charges, and lepton and kaon particle identification techniques.
  The Same Side information must necessarily exclude \Bs~decay
  products, attempting to build discriminants based upon 
  fragmentation tracks; identified kaons from fragmentation 
  (produced in conjunction with the \Bs) provide powerful
  tagging information.  Again, a variety of kinematic and
  particle-identification-based discriminants are used.
  The Same and Opposite Side tags are then combined to yield an
  overall Initial State tag.  The new ALEPH analyses~\cite{ALEPH} use series of
  neural networks (NN) to combine information, the final NN output is
  shown in Figure~\ref{fig:alephinitial}.
\begin{figure}[htb]
\begin{center}
\epsfig{file=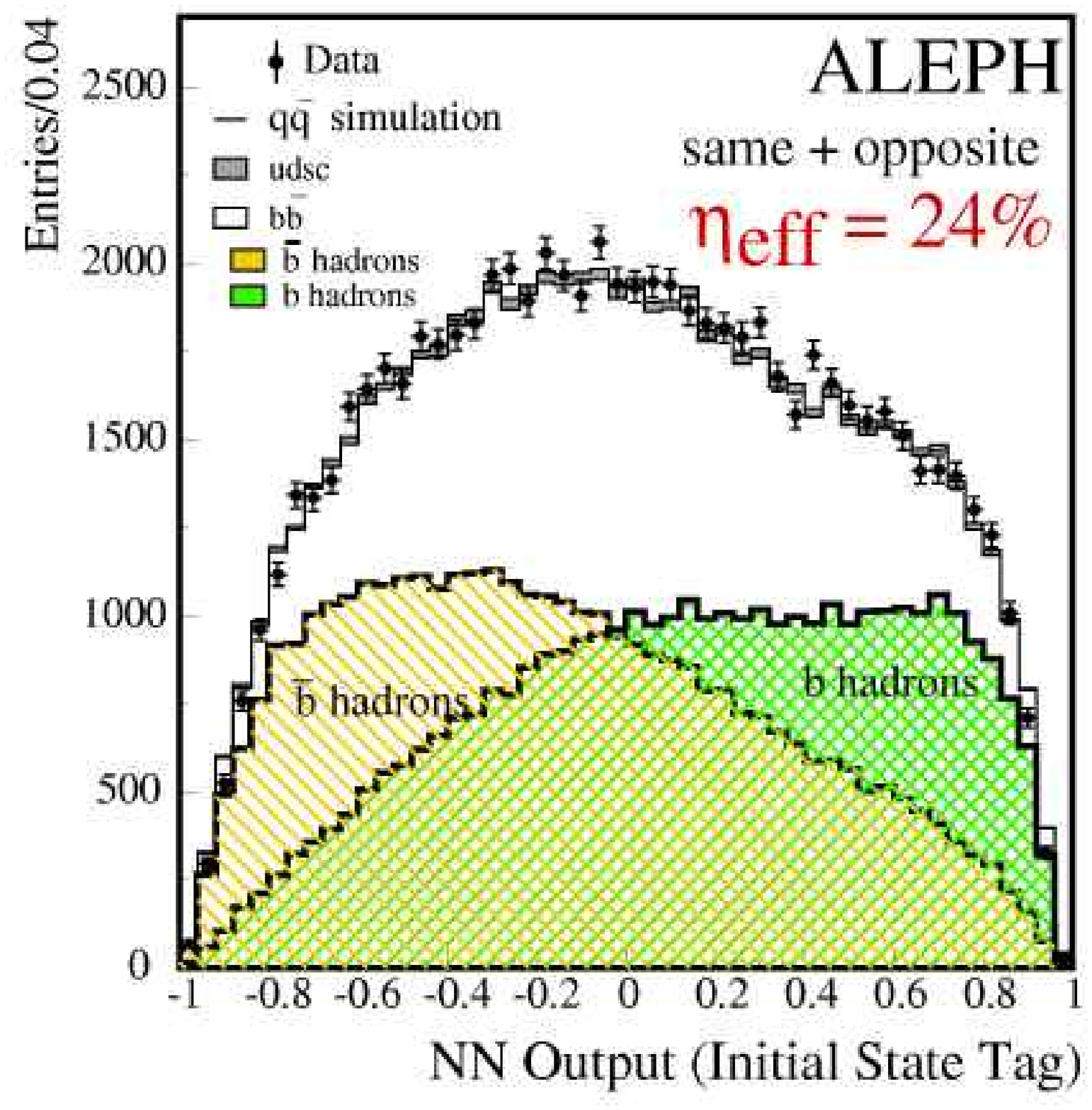,height=7cm}
\caption{{\footnotesize The ALEPH neural network-based Initial State
tagging variable shown here for the semi-inclusive analysis in selected
data (dots with error bars) and simulation (histograms)~\cite{ALEPH}.}}
\label{fig:alephinitial}
\end{center}
\end{figure}
    
\item{{\bf Measurement of Proper Time of the \Bs~Decay}} \\
  Proper time $t$ is given as $t=lm/p$ where $m$ is the \Bs~ mass;
  the \Bs~momentum $p$ and the measured decay length $l$ must be measured.
  For the fully exclusive mode, the \Bs~momentum is determined with
  excellent precision from knowledge of the momenta of all of the
  decay products.  In more inclusive selections with semileptonic
  \Bs~decays, a correction is done to account for missing neutrino momentum based upon
  event energy-momentum conservation; uncertainties associated with
  this correction procedure dominate the momentum resolution.  The
  decay length is determined by the distance between the primary
  vertex and the \Bs~decay vertex.  At LEP, the primary vertex may be 
  determined on an event-by-event basis.  The secondary vertex
  determination is again selection-dependent, and the best precision
  ({\it e.g.}, 180\,$\mu$m for the ALEPH fully exclusive
  selection~\cite{ALEPH}) is obtained from the fully exclusive modes.
  More inclusive measurements suffer from a less precise knowledge of
  the \Bs~flight direction ({\it e.g.}, missing neutrinos from
  semileptonic B decay) onto which the primary-secondary vertex
  distance is projected.
  
\item{{\bf Determination of \dms}}\\
  A signal likelihood function can be constructed from the probability
  density function for mixed and unmixed events given in Equation~2.
  A procedure referred to as the Amplitude Method~\cite{Amplitude}
  replaces \dms~ in the probabilities by a hypothesized oscillation
  frequency $\omega$ and an {\it amplitude} $\mathcal{A}$ in front of
  the oscillation term.  This permits combination of analyses
  including the systematic uncertainties.  The likelihood is maximized
  with respect to the amplitude for each $\omega$.  An amplitude
  consistent with zero is expected for values of $\omega$ far below
  the true value of \dms; an amplitude consistent with unity is
  expected for values of $\omega$ very close to the true value of
  \dms.  A range of $\omega$ may be excluded at 95\% C.L. if
  $\mathcal{A} +1.645\sigma_{\mathcal{A}}<1$.

\end{itemize}

\section{Results}
\begin{figure}[htb]
\begin{center}
\epsfig{file=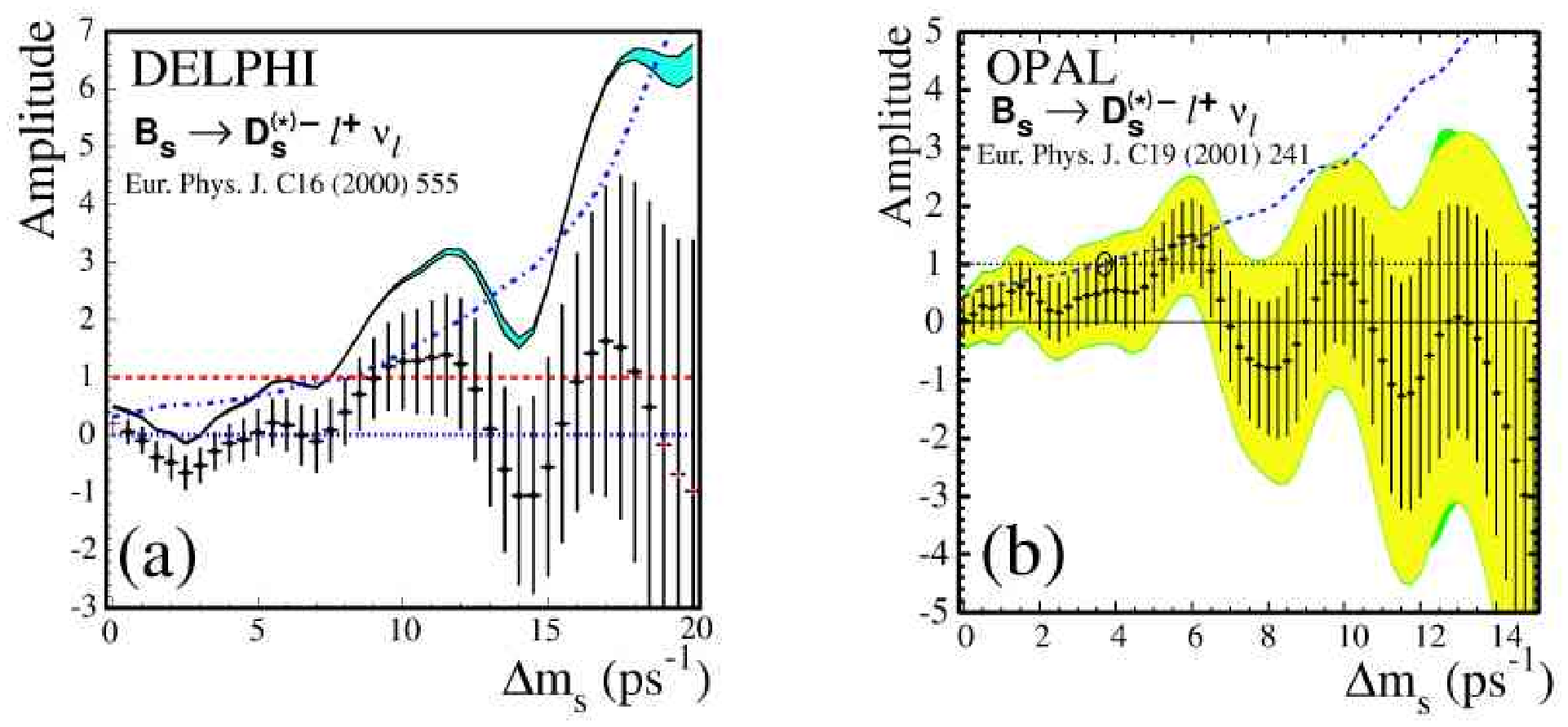,width=\textwidth}
\caption{{\footnotesize
    Plots showing Amplitude versus hypothesized \dms~for a)~the DELPHI
    semi-exclusive analysis~\cite{DELPHI}, and b)~the OPAL
    semi-exclusive analysis~\cite{OPALDsl}.  In the case of the
    a)~DELPHI analysis, an observed (expected) 95\% C.L. lower limit
    of $7.4\,\mathrm{ps}^{-1}$~($8.1\,\mathrm{ps}^{-1}$) is obtained;
    b)~OPAL analysis yields
    $1.0\,\mathrm{ps}^{-1}$~($4.1\,\mathrm{ps}^{-1}$).  }}
\label{fig:opal}
\end{center}
\end{figure}

\begin{figure}[htb]
\begin{center}
\epsfig{file=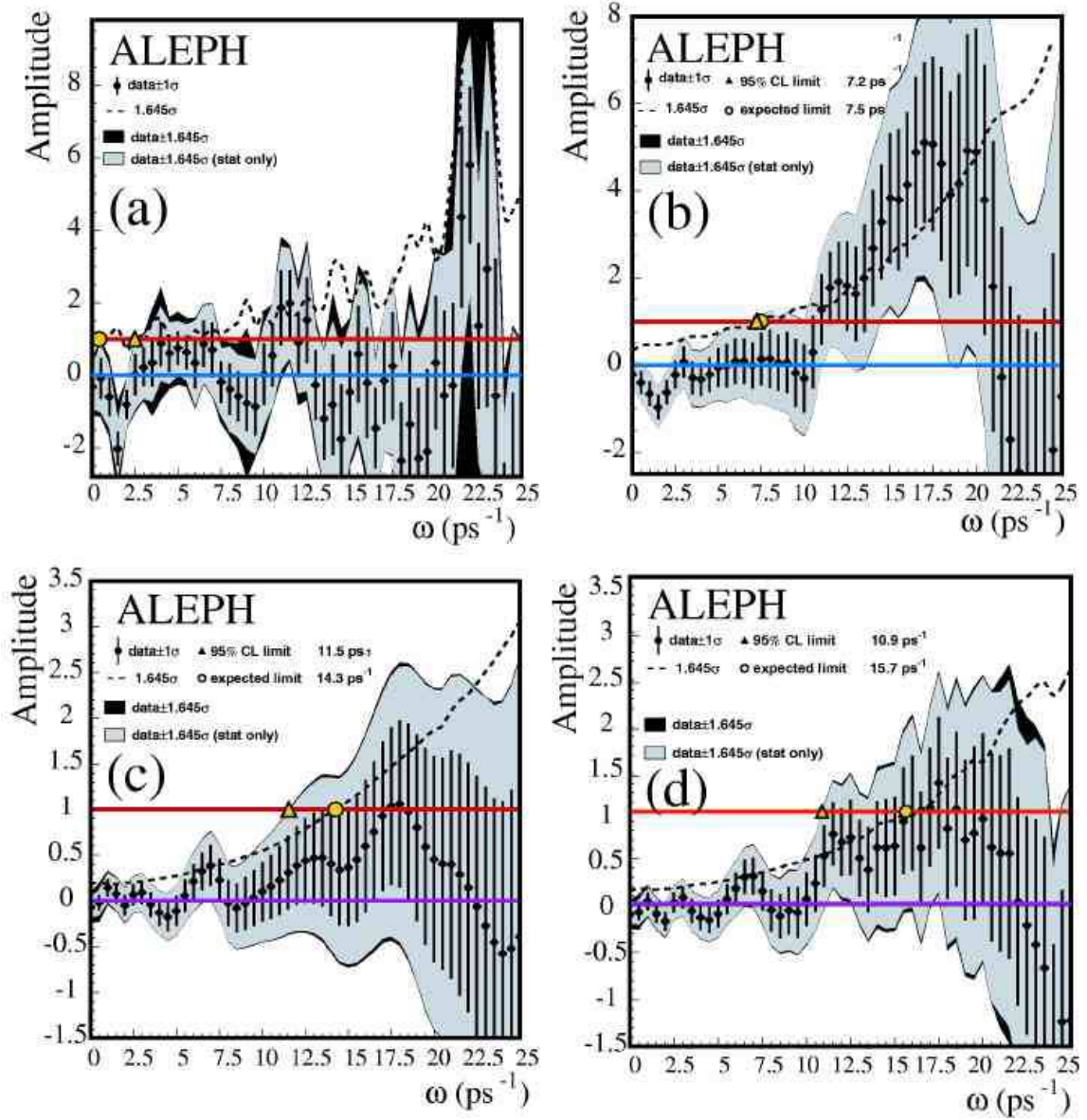,width=0.95\textwidth}
\caption{{\footnotesize The new ALEPH results~\cite{ALEPH} shown here in terms of Amplitude versus
    hypothesized \dms~($\omega$) for a) the fully exclusive analysis,
    b) the semi-exclusive D$_{\mathrm{s}}\ell$ analysis, c) the
    semi-inclusive lepton analysis, and d) the combination of the
    three.  In each case an observed (expected) 95\% C.L.  lower limit
    is set on \dms:
    a)~$2.4\,\mathrm{ps}^{-1}$~($0.3\,\mathrm{ps}^{-1}$),
    b)~$7.2\,\mathrm{ps}^{-1}$~($7.4\,\mathrm{ps}^{-1}$),
    c)~$11.4\,\mathrm{ps}^{-1}$~($14.0\,\mathrm{ps}^{-1}$), and
    d)~$10.9\,\mathrm{ps}^{-1}$~($15.7\,\mathrm{ps}^{-1}$).}}
\label{fig:aleph}
\end{center}
\end{figure}

Results of all LEP studies of \Bs-\Bsbar oscillations are summarized
in Table~\ref{tab:sum}.  Results in terms of amplitude versus
hypothesized \dms~are shown in Figure~\ref{fig:opal} for the
semi-exclusive analyses of DELPHI~\cite{DELPHI} and
OPAL~\cite{OPALDsl}.  The corresponding plots for each of the new
ALEPH analyses~\cite{ALEPH} are shown in Figure~\ref{fig:aleph}.  

\begin{figure}[htb]
\begin{center}
\epsfig{file=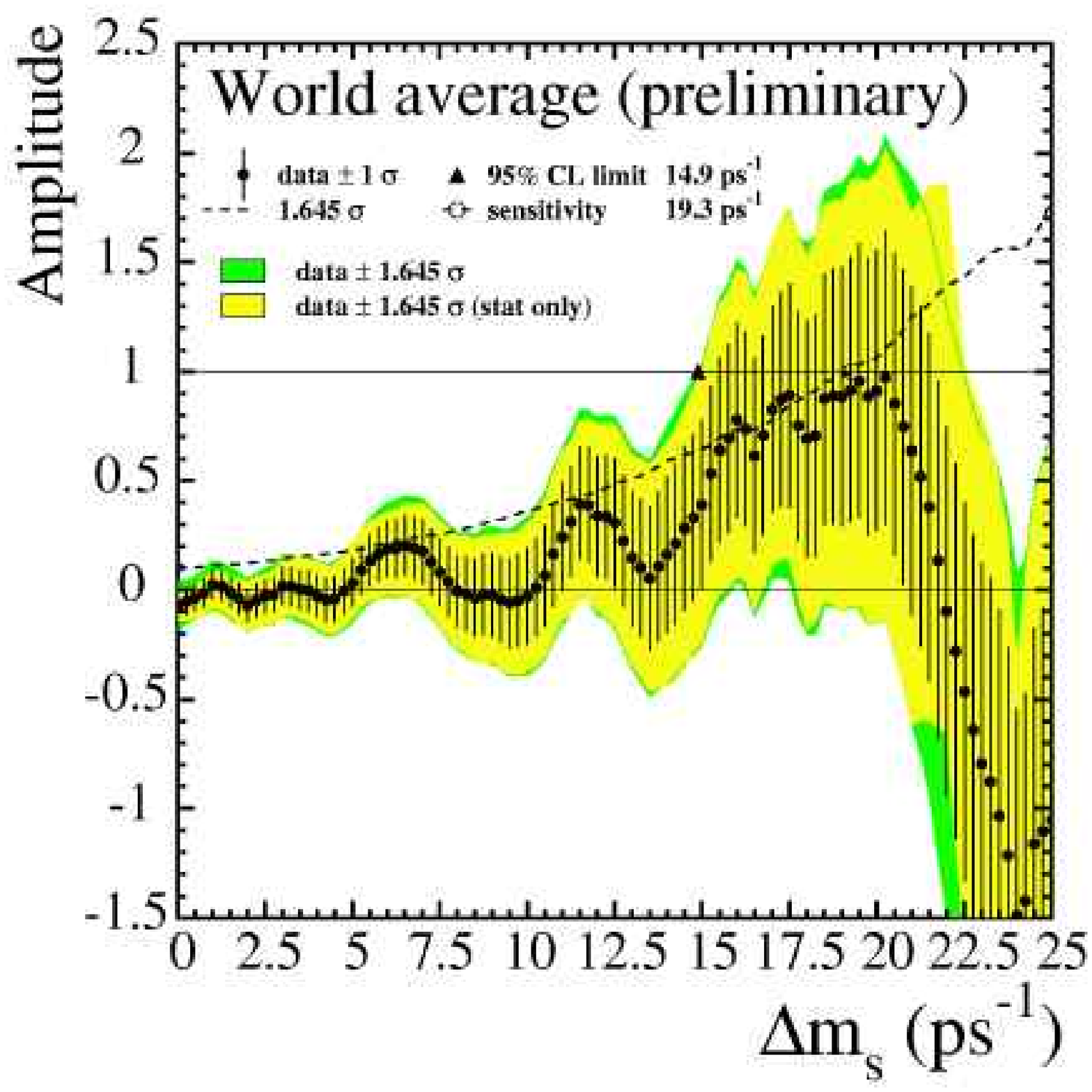,width=\textwidth}
\caption{{\footnotesize
 The combined \Bs~oscillation results from ALEPH, CDF, 
    DELPHI, OPAL, and SLD shown as amplitude versus hypothesized
    \dms~\cite{LEPBOSC}.  The dots with error bars show the fitted
    aplitude values and uncertainties.  An observed (expected) 95\%
    C.L. lower limit on \dms~of
    $14.9\,\mathrm{ps}^{-1}$~($19.3\,\mathrm{ps}^{-1}$) is obtained. }}
\label{fig:world}
\end{center}
\end{figure}

The combination of LEP results with those of CDF and SLD is shown in
Figure~\ref{fig:world}~\cite{LEPBOSC}.  For this world combination, an
observed 95\% C.L. lower limit on \dms~ of $14.9\,\mathrm{ps}^{-1}$ is
obtained with an expectation of $19.3\,\mathrm{ps}^{-1}$.
There is an apparent difference between the expected and observed
limits which suggests that a signal may lie in this region.
Furthermore, there is an enticing deviation away from consistency with
a zero amplitude hypothesis between 16 and 18\,ps$^{-1}$ which may
hint at a signal; however the statistical significance of this
deviation is below 2$\sigma$.


\section{Conclusions}
To date, no experiment has been able to resolve oscillations.  It is
presumed that the oscillation frequency lies beyond the current
experimental sensitivity to discover it at the level of 5$\sigma$;
however, there may be a hint of signal in the \dms~region between 16
and 18~ps$^{-1}$.  Further data from Run II of the Tevatron and
results of future CDF and D0 studies may soon be available with the
hope of discovering evidence for \Bs-\Bsbar~oscillations.

\section*{Acknowledgements}
This presentation summarizes the work of the ALEPH, DELPHI, and OPAL
collaborations as well as the LEP B Oscillations Working Group.  As a
member of the ALEPH collaboration, I wish to thank my colleagues in
the CERN accelerator divisions for the successful operation of LEP.  I
am grateful to Dr. Duccio Abbaneo for his help in preparing this
presentation as well as members of the ALEPH Heavy Flavour Group
and the LEP B Oscillations Working Group.  Dr. Markus Elsing
also provided valuable input and suggestions.

\end{document}